\documentclass[pdflatex,sn-nature]{sn-jnl}

\usepackage{graphicx}%
\usepackage{multirow}%
\usepackage{amsmath,amssymb,amsfonts}%
\usepackage{amsthm}%
\usepackage{mathrsfs}%
\usepackage[title]{appendix}%
\usepackage{xcolor}%
\usepackage{textcomp}%
\usepackage{manyfoot}%
\usepackage{booktabs}%
\usepackage{algorithm}%
\usepackage{algorithmicx}%
\usepackage{algpseudocode}%
\usepackage{listings}%
\usepackage[numbers]{natbib}
\usepackage{multicol}

\definecolor{red}{RGB}{255,0,0} 

\begin{document}
\newcommand{\stefania}[1]{\textcolor{red}{#1}}
\newcommand{\comment}[1]{\textcolor{blue}{#1}}
\title[Article Title]{Photo-luminescence properties of ion implanted $\text{Er}^{3+}$-defects in 4H-SiCOI towards integrated quantum photonics}


\author[1,2]{\fnm{Joshua} \sur{Bader}}
\author[3]{\fnm{Shao Qi} \sur{Lim}} 
\author[4]{\fnm{Faraz Ahmed} \sur{Inam}} 
\author[5]{\fnm{Brett C.} \sur{Johnson}} 
\author[2,6]{\fnm{Alberto} \sur{Peruzzo}} 
\author[3]{\fnm{Jeffrey} \sur{McCallum}} 
\author[7]{\fnm{Qing} \sur{Li}} 

\author[1]{\fnm{Stefania} \sur{Castelletto}} 


\affil[1]{\orgdiv{School of Engineering}, \orgname{RMIT University}, \orgaddress{\city{Melbourne}, \postcode{3000}, \state{VIC}, \country{Australia}}}
\affil[2]{\orgdiv{Quantum Photonics Laboratory and Centre for Quantum Computation and Communication Technology, School of Engineering}, \orgname{RMIT University}, \orgaddress{\city{Melbourne}, \postcode{3000}, \state{VIC}, \country{Australia}}}
\affil[3]{\orgdiv{Centre for Quantum Computation and Communication Technology, School of Physics}, \orgname{The University of Melbourne}, \orgaddress{\city{Melbourne}, \postcode{3010}, \state{VIC}, \country{Australia}}}

\affil[4]{\orgdiv{Department of Physics}, \orgname{Aligarh Muslim University}, \orgaddress{\city{Aligarh}, \postcode{202002}, \country{India}}}
\affil[5]{\orgdiv{School of Science}, \orgname{RMIT University}, \orgaddress{\city{Melbourne}, \postcode{3001 }, \country{Australia}}}

\affil[6]{\orgdiv{Advanced Research Department}, \orgname{Qubit Pharmaceuticals}, \orgaddress{\city{Paris}, \postcode{75014}, \country{France}}}
\affil[7]{\orgdiv{Electrical and Computer Engineering}, \orgname{Carnegie Mellon University}, \orgaddress{\city{Pittsburgh}, \postcode{15213}, \state{PA}, \country{USA}}}

\abstract{

Colour centres hosted in solid-state materials such as  silicon carbide and diamond are promising candidates for integration into chip-scale quantum systems. Specifically, the incorporation of these colour centres within photonic integrated circuits  may enable precise control over their inherent photo-physical properties through strong light-matter interaction. Here, we investigate ion-implanted erbium ($\text{Er}^{3+}$) defects embedded in thin-film 4H-silicon-carbide-on-insulator (4H-SiCOI). Optimized implantation conditions and thermal annealing processes designed to enhance the emission characteristics of the $\text{Er}^{3+}$-defect are reported. By examining key properties such as photoluminescence intensity, optical lifetime, and polarization, we present an analysis of ensemble $\text{Er}^{3+}$-defects within 4H-SiCOI, providing insights into their potential for future quantum applications.}




\maketitle

\section{Introduction}\label{sec1}
\begin{multicols}{2}
Silicon carbide (SiC) promises to evolve into a key platform for advancing quantum technologies. 
Over the past decade, efforts to establish SiC as a host for spin-photon-based quantum technologies have led to a notable advancement in its maturity 
\cite{lukin20204h}. Vacancy-related defects, such as the silicon vacancy \cite{liu2024silicon} and divacancy \cite{sun2023divacancy}, or rare-earth dopants \cite{astner2024vanadium, bosma2018identification} are optically active and incorporate optical spin read-out with reasonable spin coherence times; thus have attracted interest for quantum communication and computing applications as well as magnetometry \cite{niethammer2016vector}. These colour centres can act as single-photon emitters operating in the near-infrared and telecommunications band \cite{radulaski2017scalable}.
In addition to its unique advantages, SiC now offers industrial-scale wafer availability and compatibility with metal-oxide-semiconductor (CMOS) processing strategies. These features position SiC to play a key role in advancing the realization of scalable quantum computers and nuclear spin-based quantum memory architectures \cite{parthasarathy2023scalable}. Furthermore, its integration with fault-tolerant approaches \cite{ecker2024quantum} paves the way for large-scale quantum systems.

\end{multicols}
\begin{figure}[H]
    \includegraphics[width=\textwidth]{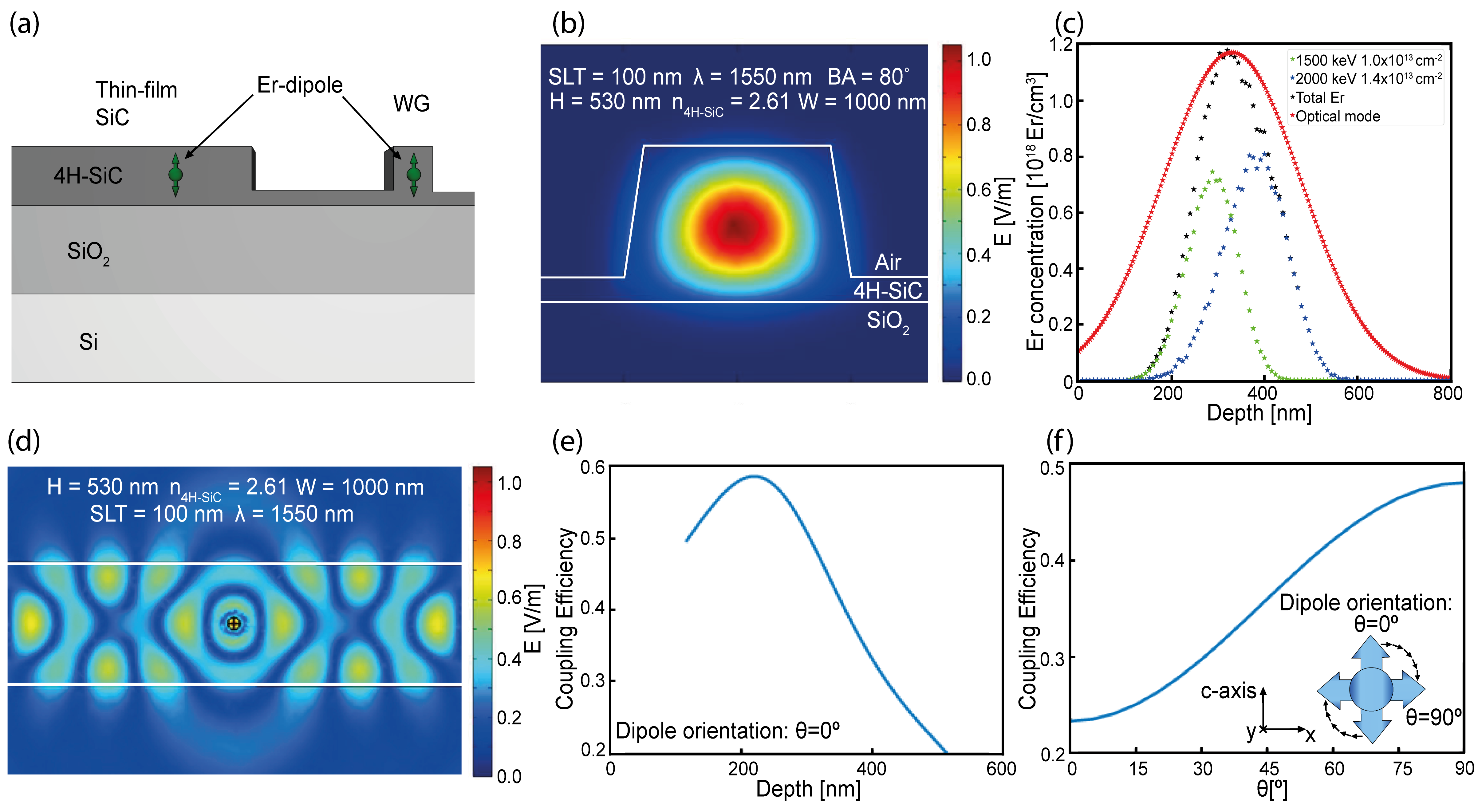}
    \caption{Waveguide mode, ion implantation profile and $\text{Er}^{3+}$-defect to waveguide coupling-simulation: a) schematic illustration of a photonic chip with two different sample architectures (SiC thin film and waveguide (WG)) ; b) ANSYS Lumerical FDTD $\text{TE}_{\text{00}}$-simulation of an air-cladded WG hosted in 4H-SiCOI with SLT: slap layer thickness, $\lambda$: light wavelength, BA: base angle of waveguide, H: waveguide height, $\text{n}_{\text{4H-SiC}}$: refractive index of 4H-SiC at desired wavelength, W: waveguide width, c) $\text{Er}^{3+}$ implantation profile in 4H-SiCOI (black), obtained from $\text{Er}^{3+}$ implantation at two different ion energies and fluences (green and blue). The optical mode is shown in red; d) simulated electric field from the vertical dipole-emission coupled to a WG with a top-view; e) simulated coupling efficiency of a vertical dipole radiating into a WG-structure versus depth of dipole location; f) simulated impact of the dipole-orientation within a SiCOI WG on the coupling efficiency for a dipole at 315 nm depth 
    with schematic photon emission direction of the dipole indicated by arrows.}
    \label{fig1}
\end{figure}
\begin{multicols}{2}

Amongst the various defects in SiC, the positively charged erbium-ion ($\text{Er}^{3+}$) is of particular interest due to its ${}^\text{4}\text{I}_{\text{13/2}}$ $\rightarrow$ ${}^\text{4}\text{I}_{\text{15/2}}$ optical transition, which has been proven to be temperature-, semiconductor- and material polytype-independent \cite{choyke1994intense}. This particular transition occurs within the $\text{4}f$-shell and is electrically shielded by its outer filled shells, which diminishes interaction with the surrounding material host \cite{babunts2000properties}. The associated electron spin is reported to be S = 1/2 \cite{baranov1997electron}.

The previously identified infrared emission with a reported zero-phonon line (ZPL) between 1528 nm and 1534 nm \cite{choyke1994intense} into the depletion minima of silica fibers/optics \cite{digonnet2001rare} is of greatest interest for quantum and classical communication applications. Specifically, this $\text{Er}^{3+}$ emission, hosted within 4H-SiCOI, may enable straightforward integration of $\text{Er}^{3+}$-based quantum emitters into low-loss existing fiber networks for long distance quantum communication \cite{stevenson2022erbium} via integrated quantum photonics.

Other novel incorporation of $\text{Er}^{3+}$-defects include Er-doped yttrium orthosilicate $\text{Y}_{2}\text{SiO}_{5}$ and yttrium oxide $\text{Y}_{2}\text{O}_{3}$ \cite{gupta2023robust} for quantum memory applications \cite{wang2023hyperfine, jiang2023quantum}.

Compared to other materials, SiC and in particular silicon-carbide-on-insulator (SiCOI) provide a quantum grade material platform, which is scalable for integrated quantum photonics in composition with $\chi^{(2)}$ and  $\chi^{(3)}$ nonlinear processes, further enhanced by a dilute nuclear spin bath \cite{Kanaigeneralised2022, seo2016quantum}. Moreover, it provides a intrinsically wide bandgap of 3.26 eV which could lead to reduced quenching of the $\text{Er}^{3+}$-photoluminescence at higher temperatures \cite{babunts2000properties}. 

$\text{Er}^{3+}$-defects have been demonstrated previously in bulk 4H-SiC \cite{choyke1994intense, parker2021infrared}, where it was shown via electron paramagnetic resonance (EPR)-spectroscopy, that implanted $\text{Er}^{3+}$-ions are most likely residing in Si-sites, thus contributing to axial C$_{3v}$-symmetry, although other hexagonal sites could also be occupied \cite{babunts2000properties}. A comparison of implanted Er-isotopes into hexagonal SiC-polytypes indicated that approximately $10\%$ of the implanted $\text{Er}^{3+}$-ions are actively involved in the optically addressable defect \cite{pasold2003erbium}.
So far, a demonstration of optically addressable $\text{Er}^{3+}$-defects in thin-film 4H-SiCOI,  which could take advantage of these unique intrinsic material capabilities, is still lacking. 

Here, we report the first realization of ensemble $\text{Er}^{3+}$-defects in thin-film 4H-SiCOI substantiated by insights into photoluminescence, optical and polarization properties. Er-ions were implanted with the goal to couple the emission of these defects to a photonic integrated circuit (PIC). In particular, coupling the optically addressable $\text{Er}^{3+}$-defect's to a waveguide mode could enable the control over the emission properties of the observed defect in a desired way.
We find a operating temperature independence from the ZPL-intensity with minor impact on the shape/linewidths of the emission. Furthermore, an optical lifetime decrease is observed within the considered 4H-SiCOI samples with polarization properties indicating a $\text{C}_{\text{3V}}$-symmetry of the $\text{Er}^{3+}$-defect.


\section{Methods}
\subsection{Material preparation}
Our 4H-SiCOI samples consist of a 630 nm thick 4H-SiC layer, bonded to a total of 2 $\mu$m Silicon-dioxide ($\text{SiO}_\text{2}$) on top of a 500 $\mu$m Si-layer handle, as shown schematically in Fig.~\ref{fig1}(a). 
The 4H-SiCOI fabrication process is described in reference \cite{CaiOctave22}. In addition, we open 100 $\mu$m wide trenches on the sample utilizing photolithography to avoid cracks during high-temperature annealing. 

\subsection{$\text{Er}^{3+}$ implantation}

We simulated the Er-ion implantation concentration depth profiles with the stopping and range of ions in matter (SRIM) software package \cite{ZIEGLER20101818}. To achieve maximal overlap with the waveguide's propagating optical mode (as shown in Fig.~\ref{fig1}(b)), two Er implantation at different ion energies and fluences were performed: (1) 1.5 MeV $1.0 \times 10^{13}$ $\text{Er}/\text{cm}^{-2}$ and (2) 2 MeV $1.4 \times 10^{13}$ $\text{Er}/\text{cm}^{-2}$. The resulting Er profile is a sum of these two profiles as shown in Fig~\ref{fig1}(c). The implantation was performed at 600\textdegree{}C under vacuum to avoid amorphization of the SiC thin film \cite{babunts2000properties}.

The simulated electric field as a result of a vertical $\text{Er}^{3+}$ dipole emitter in the waveguide is shown in Fig.~\ref{fig1}(d). The coupling efficiency between the $\text{Er}^{3+}$ emitter and waveguide was also found to depend on both, the depth location of the $\text{Er}^{3+}$ relative to the surface and angle of the $\text{Er}^{3+}$ dipole ($\theta=90^\circ$ for vertically oriented dipole along the 4H-SiC c-axis) as shown in Figs.~\ref{fig1}(e) and (f), respectively.



\subsection{$\text{Er}^{3+}$ Purcell enhancement model in thin-film and waveguide} 
All electro-dynamical calculations were carried out utilizing finite element method (FEM) based Comsol Multiphysics radio-frequency (RF) module. In these calculations, the colour-centre/emitter is considered as a radiating point dipole, which is modeled as an oscillating point current source being driven at the emission frequency $\nu=\text{c}/\lambda$ \cite{xu1999finite, novotny2012principles}. The scattering/PML boundary conditions are applied at the outer boundaries of the computational domain. The permittivity values utilized for SiC, $\text{SiO}_{2}$ and Si are based on the following reported values \cite{shaffer1971refractive, green2008self}.

The influence of the local environment can be fully expressed in terms of the classical local density of optical states (LDOS) model \cite{xu1999finite}. The dipole’s spontaneous emission rates relative to a reference system are known to be exactly the same under both classical and quantum treatments \cite{xu1999finite}. 
In our calculations which are based on the classical electro-dynamical treatment of the emitter as a radiating point dipole, the total power radiated by the dipole is calculated over a closed surface enclosing the point dipole emitter. The relative decay rate $\gamma$ is then calculated as $\gamma= \text{P}/\text{P}_{r}$, where $\text{P}_{r}$ is the power corresponding to the reference system. Here, the reference system is bulk SiC.

The coupling efficiency is calculated by scaling the time-averaged power flowing through the two ends of the SiC waveguide by the total power calculated over the surface enclosing the point dipole emitter.

\subsection{Experimental setup}
Spectroscopy measurements were conducted with a custom built confocal microscope equipped with either a 785 nm or 976 nm continuous-wave laser with a subsequent selection of a suitable dichroic mirror, tailored to the applied excitation wavelength. 
We focused the excitation onto the samples with a Olympus 0.65NA 50$\times$ dry objective. The samples were placed inside a Montana cryostation operated with cycled helium to study the low temperature properties of the $\text{Er}^{3+}$-defect. In order to capture the emitted infrared photons, we employed a InGaAs avalanche-photodiode (APD) as well as a Princeton Instruments $\text{LN}_{2}$-cooled spectrometer (schematic illustration can be found in supplementary material Fig. S1). 

We added a Thorlabs MC1F2 optical beam-chopper into the excitation section of the confocal microscope, which modulated the 785 nm excitation for the optical lifetime measurement, discussed in Sec. 3.2. Within the emission-section, a 1550 $\pm$ 50 nm bandpass isolated the $\text{Er}^{3+}$-emission.

Lastly, we implemented two full polarizer's (FP) and two $\lambda$/2 waveplates to investigate the polarization of the observed defect.
\end{multicols}
\begin{figure}[H]
    \includegraphics[width=\textwidth]{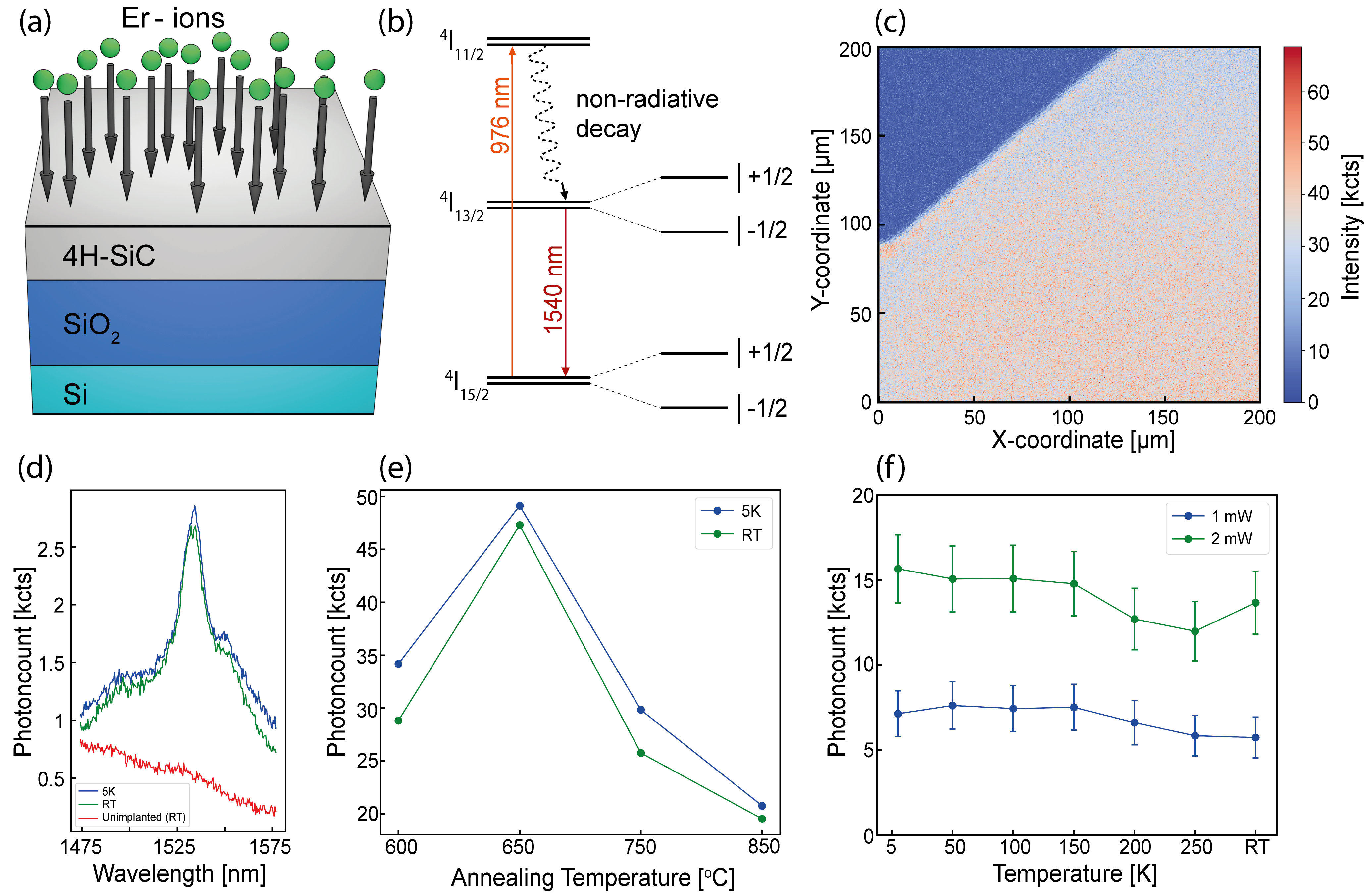}
    \caption{Ion implantation and photoluminescence measurements: a) schematic illustration of the ion implantation process into 4H-SiCOI samples, b) energy level diagram of a spin-1/2 system, c) confocal-map of a $\text{Er}^{3+}$-sample at 5 K with a 200$\times$200$ \mu$m scanning range and a resolution of 500 nm, d) spectrum of obtained $\text{Er}^{3+}$-defect's investigated measured at 5 K and RT with an additional measurement from unimplanted material at RT, e) study of ZPL intensity over annealing temperature, determined by integrating the obtained PL data-points between 1528 nm and 1534 nm, f) ZPL-intensity traces observed over various measurement-temperatures with data-points determined by integrating the obtained PL measurement data between 1528 nm and 1534 nm.} \label{fig2} 
\end{figure}

\section{Results and discussion}\label{sec2}
\begin{multicols}{2}
\subsection{Photoluminscence}
We investigated 4H-SiCOI samples with $\text{Er}^{3+}$-defects introduced via ion implantation (see Fig.~\ref{fig2}(a)). An energy level diagram for the $\text{Er}^{3+}$ ions (spin-1/2) and the relevant optical transitions are shown in Fig.~\ref{fig2}(b). The samples were imaged utilizing confocal microscopy as shown in Fig.~\ref{fig2}(c). Certain areas/sectors on the sample provided the ZPL-line shown in Fig.~\ref{fig2}(d). We conducted measurements at 5 K as well as RT and found that the ZPL-line is present and stable at both of these temperatures, in agreement with previous reports by Choyke et al. \cite{choyke1994intense}. 
We then performed an annealing study (shown in Fig.~\ref{fig2}(e)), which provides the maximum ZPL-signal at 650\textdegree{}C annealing temperature. Here, all external parameters such as the annealing atmosphere and duration were kept constant (30 minutes in Argon (Ar)).

According to the presented findings, this annealing temperature is sufficient to repair the implantation damage and to minimize other effects inducing quenching of photoluminescence.
However, the comparison between RT and 5 K revealed differences in the shape of the ZPL-peak which can be explained by utilizing the Debye-Waller Factor (DWF) \cite{debye1913interferenz} of $\approx 7.76$ $\%$ at 5 K and $\approx 8.1$ $\%$ at RT calculated as: 

\begin{equation}
\text{DWF} = \text{I}_{\text{ZPL}}/(\text{I}_{\text{ZPL}} + \text{I}_{\text{SB}})
\end{equation}

\noindent where the intensity of the ZPL ($\text{I}_{\text{ZPL}}$) was integrated from 1528 nm to 1534 nm and similarly for the sideband ($\text{I}_{\text{SB}}$) between 1450 nm to 1650 nm.
The observed temperature independence of the ZPL intensity is further illustrated in Fig.~\ref{fig2}(f) for two different excitation powers.

\subsection{Optical lifetime}
We characterized the optical lifetime of the $\text{Er}^{3+}$-defect in our thin-film 4H-SiCOI samples by analyzing the PL transient as a result of modulation of the optical excitation. To our knowledge, optical lifetimes of $\text{Er}^{3+}$ in thin-film 4H-SiCOI have 
\end{multicols}
\begin{figure}[H]
    \includegraphics[keepaspectratio, width=\textwidth]{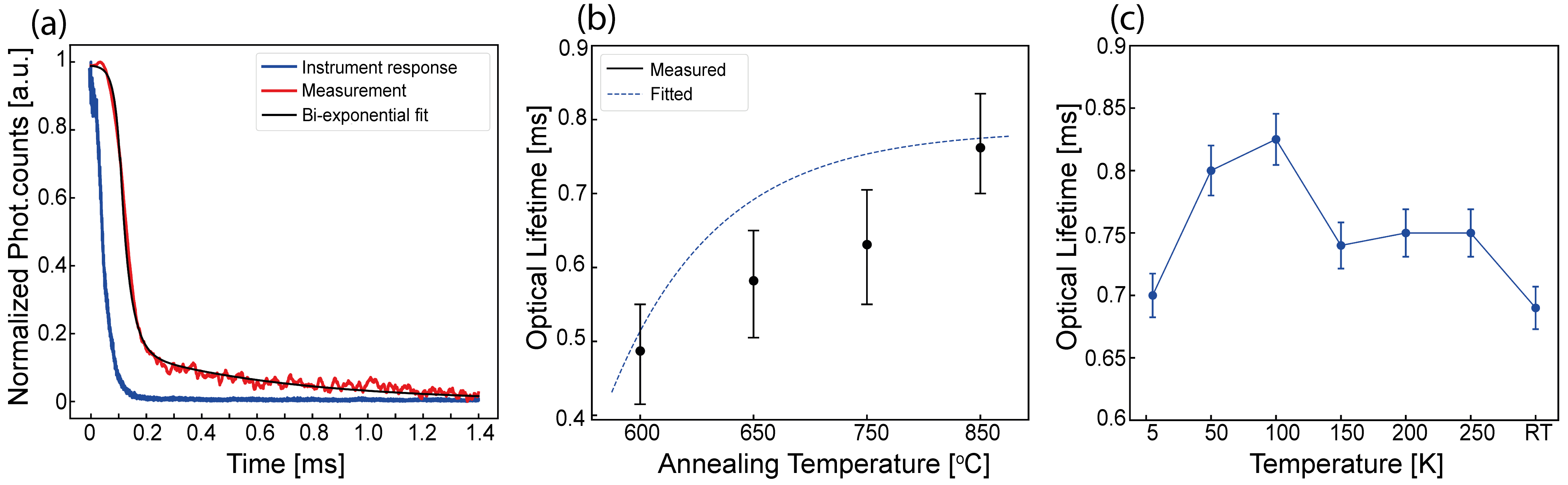}
    \centering
    \caption{\textbf{}Optical lifetime properties of observed $\text{Er}^{3+}$-defects: a) observed time-trace of an $\text{Er}^{3+}$-defect; b) Impact of annealing temperature on optical lifetime with 785 nm excitation-wavelength at RT; c) Impact of measurement temperature on optical lifetime with 785 nm excitation-wavelength.} \label{fig3}
\end{figure}
\begin{multicols}{2}
\noindent not yet been characterized. In bulk 4H-SiC, patterned lifetimes of $1.16\pm0.04$ ms as well as unpatterned lifetimes of $1.56\pm0.05$ ms have been reported \cite{parker2021infrared}. For this investigation, we modulated the excitation as elaborated in Sec. 2.4 and synchronized the detector output with the modulation trigger.
Both, the instrument response and $\text{Er}^{3+}$ emission transients were recorded. A bi-exponential fit based on 

\begin{equation}
    \tau_{\text{Fit}} = \text{a} \cdot e^{-\frac{\text{b}}{\tau_1}} + \text{c} \cdot e^{-\frac{\text{d}}{\tau_2}}
\end{equation}

\noindent was performed on the $\text{Er}^{3+}$ emission component of the transient. This gave optical lifetimes of $\tau_1 \approx 61\pm 4$ $\mu s$ and $\tau_2 \approx 582\pm 30$ $\mu s$ as illustrated in Fig.~\ref{fig3}(a). 

We performed optical lifetime measurements on five different $\text{Er}^{3+}$-defects for each performed annealing step, as illustrated in Fig.~\ref{fig3}(b). We subsequently averaged the obtained $\tau_2$-values and observed an overall lifetime variation of approximately 150 $\mu$s for each annealing step with the shortest averaged lifetime provided by the unannealed sample at 487 $\pm$ 38 $\mu$s. Furthermore, we identify an increase of the observed lifetime with higher annealing temperature in agreement with previous reports \cite{sullivan2023quasi} with average optical lifetimes determined at 582 $\pm$ 72 $\mu$s , 631 $\pm$ 77 $\mu$s as well as 762 $\pm$ 89 $\mu$s for 650\textdegree{}C, 750\textdegree{}C and 850\textdegree{}C respectively. 
Moreover, we fitted the data with the exponential fit based on

\begin{equation}
     \tau_{\text{Fit}} = \tau_{\text{max}} \cdot (1 - e^{-\frac{\text{T}}{\tau_{\text{meas}}}}) 
\end{equation}

to extrapolate the maximum achievable optical lifetime $\tau_{\text{max}}$ which has been determined to 786 $\pm$ 62 $\mu$s, with T as fitting parameter.

We investigated the temperature dependence of the optical lifetime  from 5 K to RT as shown in Fig.~\ref{fig3}(c) by conducting several measurements over a temperature range from 5 K to RT and observed an overall variation of 135 $\mu$s with the maximum lifetime observed at 100 K of 825 $\pm$ 42 $\mu$s.
We identified that the sample annealed at the highest temperature produced the longest observed lifetime of 762 $\pm$ 89 $\mu$s within the 4H-SiCOI.

\begin{figure}[H]
    \centering
    \includegraphics[keepaspectratio, width=0.7\columnwidth]{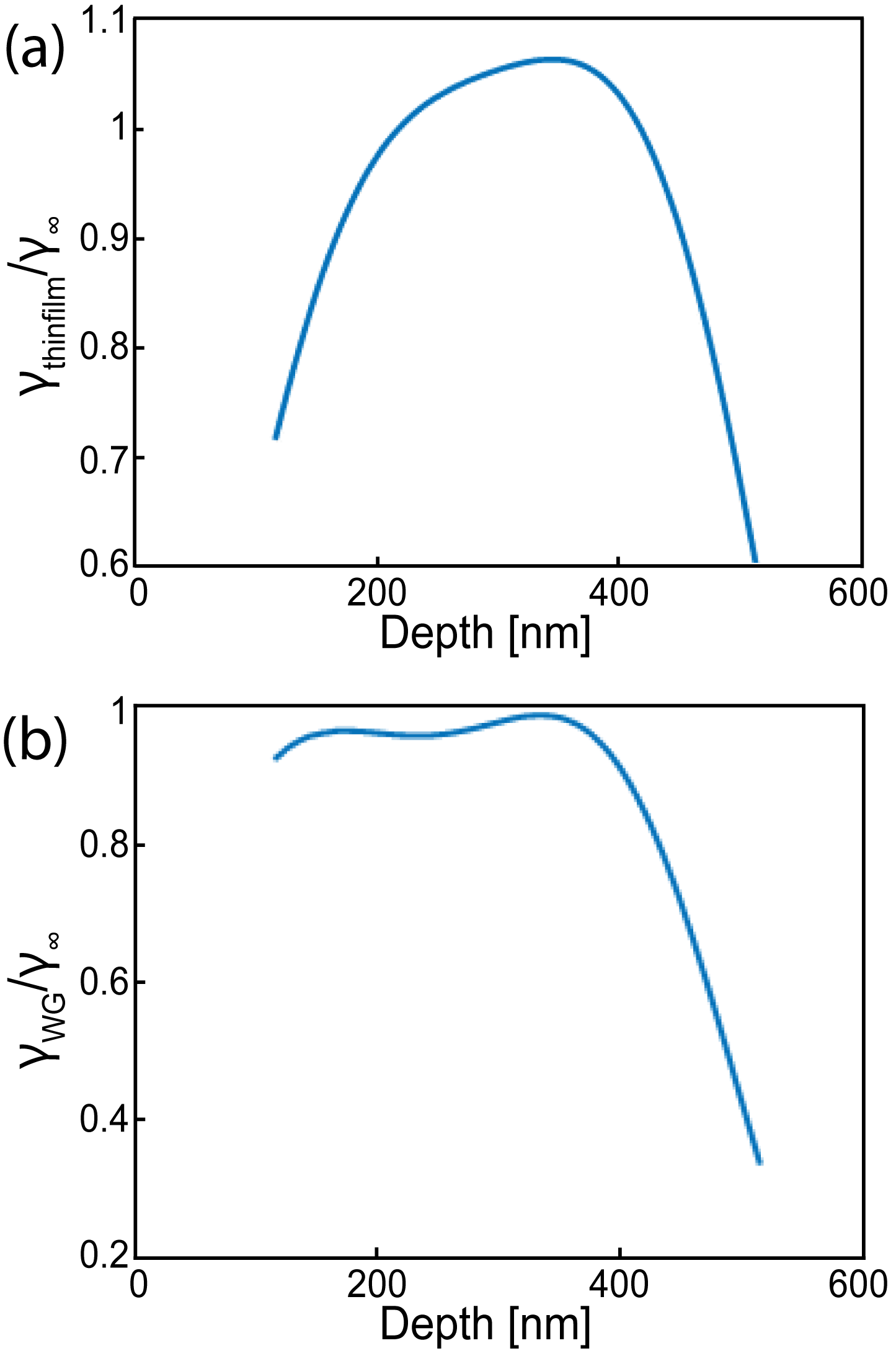}
    \caption{Dipole-interaction simulation within thin-film 4H-SiC or WG hosted in 4H-SiCOI: a) decay rate simulation of $\text{Er}^{3+}$-dipoles within thin-film 4H-SiC relative to bulk ($\gamma_{\infty}$); b) decay rate simulation of $\text{Er}^{3+}$-dipole within 4H-SiC waveguides relative to bulk ($\gamma_{\infty}$)} \label{fig4}
\end{figure}

By simulating a dipole vertically oriented in thin film 4H-SiC at various depths, we determine a Purcell's enhancement, $\gamma_{thin film}/\gamma_{\infty}$ from 0.8 to 1.1, as illustrated in Fig.~\ref{fig4}(a), with a maximum around 50 nm just below the centre of the material.
Conversely within a  waveguide-structure, we do not expect a variation of the Purcell's enhancement for dipole location below the centre of the structure, while Purcell's enhancement is suppressed for dipoles close to the waveguide surface, see Fig.~\ref{fig4}(b). This can arise due to small electric field confinements at the WG/thin-film interface which could be caused by the considered structure.



Furthermore, geometry-variations  of the thin film-layer and WG can introduce significant modifications in the local density of states, as observed with $\text{Er}^{3+}$ doped Silicon on Insulator \cite{gritsch2022narrow}. Here, other relevant observed lifetime reductions could be caused by the generally rearranged crystallinity of the thin-film 4H-SiC-layer which cannot be completely healed from implantation-damage. This could have resulted in additional defects generated by implantation damage providing non-radiative decay pathways. Therefore, the observed faster decay, described above, could be attributed to non-radiative effects as the lifetime increases with annealing temperature with approximately 10 $\%$ lifetime reduction caused by the thin film layer as well as ensemble photon-emitter properties. 

\subsection{Absorption and emission polarization }\label{sec3}

\begin{figure}[H] 
        \centering

        \includegraphics[keepaspectratio, width=0.7\columnwidth]{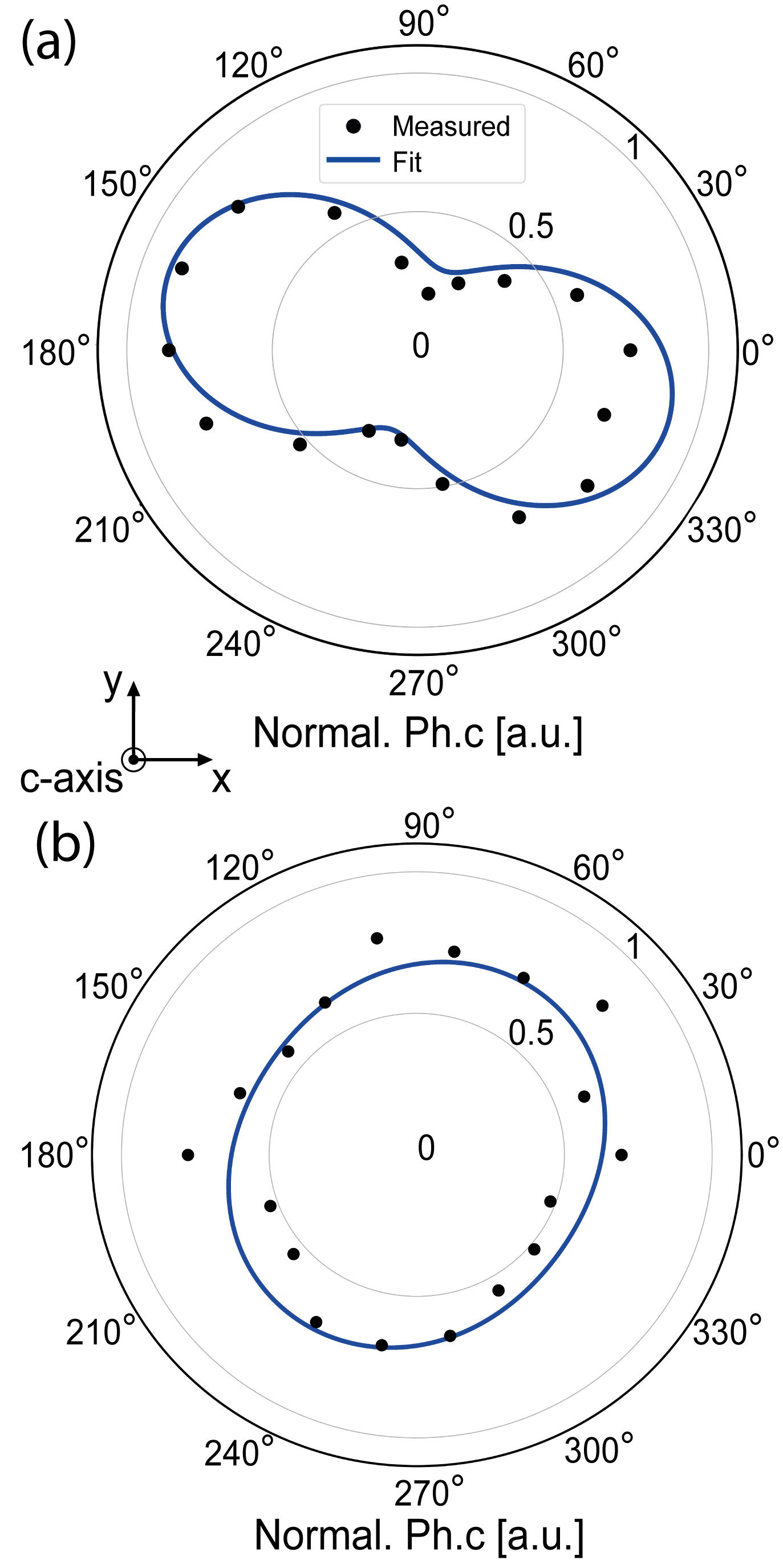}    
        \caption{Polarization properties of observed $\text{Er}^{3+}$-defects: a) observed photon absorption dipole; b) observed photon emission properties from the ensemble-defect.}
    \label{fig5}
\end{figure}

We observed an absorption dipole by rotating a 780 nm HWP in combination with an FP in the excitation section of the experimental setup with a partial polarization ratio $P_{\text{R}}$ of 0.38 (as shown in Fig.~\ref{fig5}(a)), calculated as \cite{hecht2012optics}
\begin{equation}
    I(\theta_{\text{HWP}}) = I_{\text{0}}\cdot cos^2(2\cdot \theta_{\text{HWP}}) + I_{\text{unpolarized}}
\end{equation}
where $I_{\text{0}}$ and $I_{\text{unpolarized}}$ are the polarized and unpolarized light intensities, respectively.
That subsequently led to the determination of the polarization ratio $P_{\text{R}}$, calculated as
\begin{equation}
     P_{\text{R}} = I_{\text{0}}/(I_{\text{0}}+I_{\text{unpolarized}}) 
\end{equation}
In terms of the emission polarization, we observed unpolarized dipole behaviour (see Fig.~\ref{fig5}(b)) by rotating a 1550 nm HWP in combination with an FP in the emission section of the confocal microscope with the absorption 780 nm HWP set to its maximum. A polarization ratio of $\approx$ 0.21 is determined with Eq. (4) and (5). The fit for both measurements is calculated as 
\begin{equation}
     I = a + b \cdot sin^2(\theta_{\text{HWP}}+\phi)
\end{equation}

with $a, b$ and $\phi$ as fit parameter \cite{wang2020experimental}. 
The lack of polarization in emission could be attributed to the presence of a non radiative metastable state \cite{berhane2018photophysics}. 
This indicates an orientation-offset between the absorption and emission dipole which could point to a $\text{C}_{\text{3V}}$-symmetry of the observed defects with a dominant alignment from the absorption dipole with the c-axis of the crystal.  

In particular, the dipole offset could be caused by spin-orbit coupling where the intrinsic electron spin interacts with orbital states \cite{dong2019spin}
as well as non-radiative relaxation which could be caused by vibrational modes within the SiC-lattice \cite{kenyon2002recent, miniscalco1991erbium}.
Moreover, the excited ${}^\text{4}\text{I}_{\text{13/2}}$-state could also transition through different Stark levels before relaxing towards the ${}^\text{4}\text{I}_{\text{15/2}}$ ground-state which could contribute to different polarization characteristics, as observed here \cite{digonnet2001rare}.





\section{Conclusions and Outlook }\label{sec4}
Overall, we successfully demonstrated the characterization of ensemble $\text{Er}^{3+}$-defects embedded in thin-film 4H-SiCOI. We provided insights in the photoluminescence, optical and polarization properties. With a strong observed independence from the operating temperature, this defect could be further investigated on a single-photon level at room temperature with the proposed annealing strategy applied. The observed optical lifetime is shorter in thin-film 4H-SiC layers compared to bulk material, which verified conducted simulations. The $\text{C}_{\text{3V}}$-symmetry from the observed defect deems it suitable for application in integrated quantum photonics which can enhance certain favorable characteristics, as previously demonstrated in bulk SiC \cite{radulaski2017scalable, babin2022fabrication}. Additionally, this specific combination could also lead to the composition of nuclear spins with the electron spin associated with the paramagnetic defect, which can enable the operation of quantum gates \cite{babin2022fabrication} as well as quantum memory capabilities \cite{parthasarathy2023scalable, bourassa2020entanglement}. This capability facilitates the information-transfer on chip via entanglement distribution \cite{Bhaskar2020}.


Furthermore, a recent work \cite{ramirez2024integrated} studied theoretically the utilization of $\text{Er}^{3+}$-defects embedded into ring resonators as integrated photon pairs source generated by spontaneous four-wave mixing (SFWM) within 4H-SiCOI, where substantial advantages like low two-photon absorption and free carrier absorption which negatively affect the performance, come into consideration \cite{xing2020high, xing2019cmos}. This proposed design influences the spectral response to be compatible with quantum memories hosted in a $\text{Er}^{3+}$:$\text{Y}_{2}\text{SiO}_{5}$ matrix.

The presented study points towards a new frontier where the composition of radiative defect within the telecommunication band and integrated quantum photonics play a crucial role. Especially, 4H-SiCOI has potentially a bright future with its intrinsic material capabilities, like $\chi^{(2)}$ and  $\chi^{(3)}$ processes with dilute nuclear spin bath in a quantum grade, CMOS-compatible matrix.

\backmatter

\section*{Acknowledgements}

J.McC. acknowledges the Australian Government Australian Research Council under the Centre of Excellence scheme (No: CE170100012). 

Q. Li is supported by the National Science Foundation of the United States of America under Grant No. 2240420.

The optical confocal characterizations have been conducted in the RMIT laboratories, partially funded by the ARC Centre of Excellence for Nanoscale BioPhotonics (No. CE140100003), and the LIEF scheme grant (No. LE140100131). 

We acknowledge the use of the NCRIS Heavy Ion Accelerator platform (HIA) for access and support to the ion implantation equipment at the Australian National University.

This work was performed in part at the RMIT Micro Nano Research Facility (MNRF) in the Victorian Node of the Australian National Fabrication Facility (ANFF).

All authors acknowledge the work from Ruixuan Wang and Jingwei Li, who contributed to the fabrication of the investigated samples. 

\section*{Authors contribution}

S.C., A.P., J.B., J.McC., S.Q.L.: Conceptualization, experimental design and methodology. Q.L., S.Q.L., J.McC.: Fabrication and ion implantation. J.B., S.C., F.I., B.C.J.: Data acquisition, modeling, curation, visualization, and investigation. S.C., A.P.: Supervision. J.B, S.C.:
Writing original manuscript draft, reviewing and editing. All authors contributed to writing and commenting the final manuscript.

\section*{Competing Interests}
The authors have no competing interests

\section*{Data availability}
The presented data is available upon request from the corresponding author.

\end{multicols}


\bibliography{sn-bibliography}

\begin{thebibliography}{10}
\expandafter\ifx\csname url\endcsname\relax
  \def\url#1{\burl{#1}}\fi
\expandafter\ifx\csname urlprefix\endcsname\relax\def\urlprefix{URL }\fi
\providecommand{\bibinfo}[2]{#2}
\providecommand{\eprint}[2][]{\url{#2}}
\providecommand{\doi}[1]{\url{https://doi.org/#1}}
\bibcommenthead

\bibitem{lukin20204h}
\bibinfo{author}{Lukin, D.~M.} \emph{et~al.}
\newblock \bibinfo{title}{4h-silicon-carbide-on-insulator for integrated quantum and nonlinear photonics}.
\newblock \emph{\bibinfo{journal}{Nature Photonics}} \textbf{\bibinfo{volume}{14}}, \bibinfo{pages}{330--334} (\bibinfo{year}{2020}).

\bibitem{liu2024silicon}
\bibinfo{author}{Liu, D.} \emph{et~al.}
\newblock \bibinfo{title}{The silicon vacancy centers in sic: determination of intrinsic spin dynamics for integrated quantum photonics}.
\newblock \emph{\bibinfo{journal}{npj Quantum Information}} \textbf{\bibinfo{volume}{10}}, \bibinfo{pages}{72} (\bibinfo{year}{2024}).

\bibitem{sun2023divacancy}
\bibinfo{author}{Sun, T.} \emph{et~al.}
\newblock \bibinfo{title}{Divacancy and silicon vacancy color centers in 4h-sic fabricated by hydrogen and dual ions implantation and annealing}.
\newblock \emph{\bibinfo{journal}{Ceramics International}} \textbf{\bibinfo{volume}{49}}, \bibinfo{pages}{7452--7465} (\bibinfo{year}{2023}).

\bibitem{astner2024vanadium}
\bibinfo{author}{Astner, T.} \emph{et~al.}
\newblock \bibinfo{title}{Vanadium in silicon carbide: Telecom-ready spin centres with long relaxation lifetimes and hyperfine-resolved optical transitions}.
\newblock \emph{\bibinfo{journal}{Quantum Science and Technology}} \textbf{\bibinfo{volume}{9}}, \bibinfo{pages}{035038} (\bibinfo{year}{2024}).

\bibitem{bosma2018identification}
\bibinfo{author}{Bosma, T.} \emph{et~al.}
\newblock \bibinfo{title}{Identification and tunable optical coherent control of transition-metal spins in silicon carbide}.
\newblock \emph{\bibinfo{journal}{npj Quantum Information}} \textbf{\bibinfo{volume}{4}}, \bibinfo{pages}{48} (\bibinfo{year}{2018}).

\bibitem{niethammer2016vector}
\bibinfo{author}{Niethammer, M.} \emph{et~al.}
\newblock \bibinfo{title}{Vector magnetometry using silicon vacancies in 4 h-sic under ambient conditions}.
\newblock \emph{\bibinfo{journal}{Physical Review Applied}} \textbf{\bibinfo{volume}{6}}, \bibinfo{pages}{034001} (\bibinfo{year}{2016}).

\bibitem{radulaski2017scalable}
\bibinfo{author}{Radulaski, M.} \emph{et~al.}
\newblock \bibinfo{title}{Scalable quantum photonics with single color centers in silicon carbide}.
\newblock \emph{\bibinfo{journal}{Nano letters}} \textbf{\bibinfo{volume}{17}}, \bibinfo{pages}{1782--1786} (\bibinfo{year}{2017}).

\bibitem{parthasarathy2023scalable}
\bibinfo{author}{Parthasarathy, S.~K.} \emph{et~al.}
\newblock \bibinfo{title}{Scalable quantum memory nodes using nuclear spins in silicon carbide}.
\newblock \emph{\bibinfo{journal}{Physical Review Applied}} \textbf{\bibinfo{volume}{19}}, \bibinfo{pages}{034026} (\bibinfo{year}{2023}).

\bibitem{ecker2024quantum}
\bibinfo{author}{Ecker, S.} \emph{et~al.}
\newblock \bibinfo{title}{Quantum communication networks with defects in silicon carbide}.
\newblock \emph{\bibinfo{journal}{arXiv preprint arXiv:2403.03284}}  (\bibinfo{year}{2024}).

\bibitem{choyke1994intense}
\bibinfo{author}{Choyke, W.} \emph{et~al.}
\newblock \bibinfo{title}{Intense erbium-1.54-$\mu$m photoluminescence from 2 to 525 k in ion-implanted 4h, 6h, 15r, and 3c sic}.
\newblock \emph{\bibinfo{journal}{Applied physics letters}} \textbf{\bibinfo{volume}{65}}, \bibinfo{pages}{1668--1670} (\bibinfo{year}{1994}).

\bibitem{babunts2000properties}
\bibinfo{author}{Babunts, R.} \emph{et~al.}
\newblock \bibinfo{title}{Properties of erbium luminescence in bulk crystals of silicon carbide}.
\newblock \emph{\bibinfo{journal}{Physics of the Solid State}} \textbf{\bibinfo{volume}{42}}, \bibinfo{pages}{829--835} (\bibinfo{year}{2000}).

\bibitem{baranov1997electron}
\bibinfo{author}{Baranov, P.}, \bibinfo{author}{Ilyin, I.~V.} \& \bibinfo{author}{Mokhov, E.}
\newblock \bibinfo{title}{Electron paramagnetic resonance of erbium in bulk silicon carbide crystals}.
\newblock \emph{\bibinfo{journal}{Materials Science Forum}} \textbf{\bibinfo{volume}{258}}, \bibinfo{pages}{1539--1544} (\bibinfo{year}{1997}).

\bibitem{digonnet2001rare}
\bibinfo{author}{Digonnet, M.~J.}
\newblock \emph{\bibinfo{title}{Rare-earth-doped fiber lasers and amplifiers, revised and expanded}}  (\bibinfo{publisher}{CRC press}, \bibinfo{year}{2001}).

\bibitem{stevenson2022erbium}
\bibinfo{author}{Stevenson, P.} \emph{et~al.}
\newblock \bibinfo{title}{Erbium-implanted materials for quantum communication applications}.
\newblock \emph{\bibinfo{journal}{Physical Review B}} \textbf{\bibinfo{volume}{105}}, \bibinfo{pages}{224106} (\bibinfo{year}{2022}).

\bibitem{gupta2023robust}
\bibinfo{author}{Gupta, S.}, \bibinfo{author}{Wu, X.}, \bibinfo{author}{Zhang, H.}, \bibinfo{author}{Yang, J.} \& \bibinfo{author}{Zhong, T.}
\newblock \bibinfo{title}{Robust millisecond coherence times of erbium electron spins}.
\newblock \emph{\bibinfo{journal}{Physical Review Applied}} \textbf{\bibinfo{volume}{19}}, \bibinfo{pages}{044029} (\bibinfo{year}{2023}).

\bibitem{wang2023hyperfine}
\bibinfo{author}{Wang, S.-J.}, \bibinfo{author}{Chen, Y.-H.}, \bibinfo{author}{Longdell, J.~J.} \& \bibinfo{author}{Zhang, X.}
\newblock \bibinfo{title}{Hyperfine states of erbium doped yttrium orthosilicate for long-coherence-time quantum memories}.
\newblock \emph{\bibinfo{journal}{Journal of Luminescence}} \textbf{\bibinfo{volume}{262}}, \bibinfo{pages}{119935} (\bibinfo{year}{2023}).

\bibitem{jiang2023quantum}
\bibinfo{author}{Jiang, M.-H.} \emph{et~al.}
\newblock \bibinfo{title}{Quantum storage of entangled photons at telecom wavelengths in a crystal}.
\newblock \emph{\bibinfo{journal}{Nature Communications}} \textbf{\bibinfo{volume}{14}}, \bibinfo{pages}{6995} (\bibinfo{year}{2023}).

\bibitem{Kanaigeneralised2022}
\bibinfo{author}{Kanai, S.} \emph{et~al.}
\newblock \bibinfo{title}{Generalized scaling of spin qubit coherence in over 12,000 host materials}.
\newblock \emph{\bibinfo{journal}{Proceedings of the National Academy of Sciences}} \textbf{\bibinfo{volume}{119}} (\bibinfo{year}{2022}).

\bibitem{seo2016quantum}
\bibinfo{author}{Seo, H.} \emph{et~al.}
\newblock \bibinfo{title}{Quantum decoherence dynamics of divacancy spins in silicon carbide}.
\newblock \emph{\bibinfo{journal}{Nature communications}} \textbf{\bibinfo{volume}{7}}, \bibinfo{pages}{12935} (\bibinfo{year}{2016}).

\bibitem{parker2021infrared}
\bibinfo{author}{Parker, R.} \emph{et~al.}
\newblock \bibinfo{title}{Infrared erbium photoluminescence enhancement in silicon carbide nano-pillars}.
\newblock \emph{\bibinfo{journal}{Journal of Applied Physics}} \textbf{\bibinfo{volume}{130}} (\bibinfo{year}{2021}).

\bibitem{pasold2003erbium}
\bibinfo{author}{Pasold, G.} \emph{et~al.}
\newblock \bibinfo{title}{Erbium-related band gap states in 4{H}- and 6{H}-silicon carbide}.
\newblock \emph{\bibinfo{journal}{Journal of applied physics}} \textbf{\bibinfo{volume}{93}}, \bibinfo{pages}{2289--2291} (\bibinfo{year}{2003}).

\bibitem{CaiOctave22}
\bibinfo{author}{Cai, L.}, \bibinfo{author}{Li, J.}, \bibinfo{author}{Wang, R.} \& \bibinfo{author}{Li, Q.}
\newblock \bibinfo{title}{Octave-spanning microcomb generation in 4h-silicon-carbide-on-insulator photonics platform}.
\newblock \emph{\bibinfo{journal}{Photon. Res.}} \textbf{\bibinfo{volume}{10}}, \bibinfo{pages}{870--876} (\bibinfo{year}{2022}).

\bibitem{ZIEGLER20101818}
\bibinfo{author}{Ziegler, J.~F.}, \bibinfo{author}{Ziegler, M.} \& \bibinfo{author}{Biersack, J.}
\newblock \bibinfo{title}{Srim – the stopping and range of ions in matter (2010)}.
\newblock \emph{\bibinfo{journal}{Nuclear Instruments and Methods in Physics Research Section B: Beam Interactions with Materials and Atoms}} \textbf{\bibinfo{volume}{268}}, \bibinfo{pages}{1818--1823} (\bibinfo{year}{2010}).

\bibitem{xu1999finite}
\bibinfo{author}{Xu, Y.} \emph{et~al.}
\newblock \bibinfo{title}{Finite-difference time-domain calculation of spontaneous emission lifetime in a microcavity}.
\newblock \emph{\bibinfo{journal}{JOSA B}} \textbf{\bibinfo{volume}{16}}, \bibinfo{pages}{465--474} (\bibinfo{year}{1999}).

\bibitem{novotny2012principles}
\bibinfo{author}{Novotny, L.} \& \bibinfo{author}{Hecht, B.}
\newblock \emph{\bibinfo{title}{Principles of nano-optics}}  (\bibinfo{publisher}{Cambridge university press}, \bibinfo{year}{2012}).

\bibitem{shaffer1971refractive}
\bibinfo{author}{Shaffer, P.~T.}
\newblock \bibinfo{title}{Refractive index, dispersion, and birefringence of silicon carbide polytypes}.
\newblock \emph{\bibinfo{journal}{Applied optics}} \textbf{\bibinfo{volume}{10}}, \bibinfo{pages}{1034--1036} (\bibinfo{year}{1971}).

\bibitem{green2008self}
\bibinfo{author}{Green, M.~A.}
\newblock \bibinfo{title}{Self-consistent optical parameters of intrinsic silicon at 300 k including temperature coefficients}.
\newblock \emph{\bibinfo{journal}{Solar Energy Materials and Solar Cells}} \textbf{\bibinfo{volume}{92}}, \bibinfo{pages}{1305--1310} (\bibinfo{year}{2008}).

\bibitem{debye1913interferenz}
\bibinfo{author}{Debye, P.}
\newblock \bibinfo{title}{Interferenz von roentgenstrahlen und waermebewegung}.
\newblock \emph{\bibinfo{journal}{Annalen der Physik}} \textbf{\bibinfo{volume}{348}}, \bibinfo{pages}{49--92} (\bibinfo{year}{1913}).

\bibitem{sullivan2023quasi}
\bibinfo{author}{Sullivan, S.~E.} \emph{et~al.}
\newblock \bibinfo{title}{Quasi-deterministic localization of er emitters in thin film tio2 through submicron-scale crystalline phase control}.
\newblock \emph{\bibinfo{journal}{Applied Physics Letters}} \textbf{\bibinfo{volume}{123}} (\bibinfo{year}{2023}).

\bibitem{gritsch2022narrow}
\bibinfo{author}{Gritsch, A.}, \bibinfo{author}{Weiss, L.}, \bibinfo{author}{Fruh, J.}, \bibinfo{author}{Rinner, S.} \& \bibinfo{author}{Reiserer, A.}
\newblock \bibinfo{title}{Narrow optical transitions in erbium-implanted silicon waveguides}.
\newblock \emph{\bibinfo{journal}{Physical Review X}} \textbf{\bibinfo{volume}{12}}, \bibinfo{pages}{041009} (\bibinfo{year}{2022}).

\bibitem{hecht2012optics}
\bibinfo{author}{Hecht, E.}
\newblock \emph{\bibinfo{title}{Optics}}  (\bibinfo{publisher}{Pearson Education India}, \bibinfo{year}{2012}).

\bibitem{wang2020experimental}
\bibinfo{author}{Wang, J.-F.} \emph{et~al.}
\newblock \bibinfo{title}{Experimental optical properties of single nitrogen vacancy centers in silicon carbide at room temperature}.
\newblock \emph{\bibinfo{journal}{Acs Photonics}} \textbf{\bibinfo{volume}{7}}, \bibinfo{pages}{1611--1616} (\bibinfo{year}{2020}).

\bibitem{berhane2018photophysics}
\bibinfo{author}{Berhane, A.~M.} \emph{et~al.}
\newblock \bibinfo{title}{Photophysics of gan single-photon emitters in the visible spectral range}.
\newblock \emph{\bibinfo{journal}{Physical Review B}} \textbf{\bibinfo{volume}{97}}, \bibinfo{pages}{165202} (\bibinfo{year}{2018}).

\bibitem{dong2019spin}
\bibinfo{author}{Dong, W.}, \bibinfo{author}{Doherty, M.} \& \bibinfo{author}{Economou, S.~E.}
\newblock \bibinfo{title}{Spin polarization through intersystem crossing in the silicon vacancy of silicon carbide}.
\newblock \emph{\bibinfo{journal}{Physical Review B}} \textbf{\bibinfo{volume}{99}}, \bibinfo{pages}{184102} (\bibinfo{year}{2019}).

\bibitem{kenyon2002recent}
\bibinfo{author}{Kenyon, A.}
\newblock \bibinfo{title}{Recent developments in rare-earth doped materials for optoelectronics}.
\newblock \emph{\bibinfo{journal}{Progress in Quantum electronics}} \textbf{\bibinfo{volume}{26}}, \bibinfo{pages}{225--284} (\bibinfo{year}{2002}).

\bibitem{miniscalco1991erbium}
\bibinfo{author}{Miniscalco, W.~J.}
\newblock \bibinfo{title}{Erbium-doped glasses for fiber amplifiers at 1500 nm}.
\newblock \emph{\bibinfo{journal}{Journal of Lightwave Technology}} \textbf{\bibinfo{volume}{9}}, \bibinfo{pages}{234--250} (\bibinfo{year}{1991}).

\bibitem{babin2022fabrication}
\bibinfo{author}{Babin, C.} \emph{et~al.}
\newblock \bibinfo{title}{Fabrication and nanophotonic waveguide integration of silicon carbide colour centres with preserved spin-optical coherence}.
\newblock \emph{\bibinfo{journal}{Nature materials}} \textbf{\bibinfo{volume}{21}}, \bibinfo{pages}{67--73} (\bibinfo{year}{2022}).

\bibitem{bourassa2020entanglement}
\bibinfo{author}{Bourassa, A.} \emph{et~al.}
\newblock \bibinfo{title}{Entanglement and control of single nuclear spins in isotopically engineered silicon carbide}.
\newblock \emph{\bibinfo{journal}{Nature Materials}} \textbf{\bibinfo{volume}{19}}, \bibinfo{pages}{1319--1325} (\bibinfo{year}{2020}).

\bibitem{Bhaskar2020}
\bibinfo{author}{Bhaskar, M.} \emph{et~al.}
\newblock \bibinfo{title}{Experimental demonstration of memory-enhanced quantum communication}.
\newblock \emph{\bibinfo{journal}{Nature}} \textbf{\bibinfo{volume}{580}}, \bibinfo{pages}{1--5} (\bibinfo{year}{2020}).

\bibitem{ramirez2024integrated}
\bibinfo{author}{Ramirez, P.~T.}, \bibinfo{author}{Gomez, J.~D.}, \bibinfo{author}{Becerra, G.~R.}, \bibinfo{author}{Ramirez-Alarcon, R.} \& \bibinfo{author}{Robles, M.~G.}
\newblock \bibinfo{title}{Integrated photon pairs source in silicon carbide based on micro-ring resonators for quantum storage at telecom wavelengths}.
\newblock \emph{\bibinfo{journal}{Scientific Reports}} \textbf{\bibinfo{volume}{14}}, \bibinfo{pages}{17755} (\bibinfo{year}{2024}).

\bibitem{xing2020high}
\bibinfo{author}{Xing, P.}, \bibinfo{author}{Ma, D.}, \bibinfo{author}{Kimerling, L.~C.}, \bibinfo{author}{Agarwal, A.~M.} \& \bibinfo{author}{Tan, D.~T.}
\newblock \bibinfo{title}{High efficiency four wave mixing and optical bistability in amorphous silicon carbide ring resonators}.
\newblock \emph{\bibinfo{journal}{APL Photonics}} \textbf{\bibinfo{volume}{5}} (\bibinfo{year}{2020}).

\bibitem{xing2019cmos}
\bibinfo{author}{Xing, P.} \emph{et~al.}
\newblock \bibinfo{title}{Cmos-compatible pecvd silicon carbide platform for linear and nonlinear optics}.
\newblock \emph{\bibinfo{journal}{ACS Photonics}} \textbf{\bibinfo{volume}{6}}, \bibinfo{pages}{1162--1167} (\bibinfo{year}{2019}).

\end{thebibliography}

\end{document}